\documentclass[10pt,a4paper,onecolumn]{article}
\usepackage{marginnote}
\usepackage{graphicx}
\usepackage{xcolor}
\usepackage{authblk,etoolbox}
\usepackage{titlesec}
\usepackage{calc}
\usepackage{tikz}
\usepackage[linktocpage]{hyperref}
\hypersetup{colorlinks,breaklinks=true,
            urlcolor=[rgb]{0.0, 0.5, 1.0},
            linkcolor=[rgb]{0.0, 0.5, 1.0},
            citecolor=[rgb]{0.0, 0.5, 1.0}}
\usepackage{caption}
\usepackage{tcolorbox}
\usepackage{amssymb,amsmath}
\usepackage{ifxetex,ifluatex}
\usepackage{seqsplit}
\usepackage{xstring}

\usepackage{float}
\let\origfigure\figure
\let\endorigfigure\endfigure
\renewenvironment{figure}[1][2] {
    \expandafter\origfigure\expandafter[H]
} {
    \endorigfigure
}

\usepackage{fixltx2e} 
\usepackage[
  backend=biber,
  sorting=none
]{biblatex}
\bibliography{paper.bib}

\let\textttOrig=\texttt
\def\texttt#1{\expandafter\textttOrig{\seqsplit{#1}}}
\renewcommand{\seqinsert}{\ifmmode
  \allowbreak
  \else\penalty6000\hspace{0pt plus 0.02em}\fi}

\makeatletter
\let\href@Orig=\href
\def\href@Urllike#1#2{\href@Orig{#1}{\begingroup
    \def\Url@String{#2}\Url@FormatString
    \endgroup}}
\def\href@Notdoi#1#2{\def\tempa{#1}\def\tempb{#2}%
  \ifx\tempa\tempb\relax\href@Urllike{#1}{#2}\else
  \href@Orig{#1}{#2}\fi}
\def\href#1#2{%
  \IfBeginWith{#1}{https://doi.org}%
  {\href@Urllike{#1}{#2}}{\href@Notdoi{#1}{#2}}}
\makeatother

\newlength{\cslhangindent}
\newlength{\csllabelwidth}
  {\setlength{\parindent}{0pt}%
  \everypar{\setlength{\hangindent}{\cslhangindent}}\ignorespaces}%
  {\par}
 {
  \setlength{\parindent}{0pt}
  \ifodd #1 \everypar{\setlength{\hangindent}{\cslhangindent}}\ignorespaces\fi
  \ifnum #2 > 0
  \setlength{\parskip}{#2\baselineskip}
  \fi
 }%
 {}
\usepackage{calc}

\usepackage[top=3.5cm, bottom=3cm, right=1.5cm, left=1.0cm,
            headheight=2.2cm, reversemp, includemp, marginparwidth=4.5cm]{geometry}

\titleformat{\section}
  {\normalfont\sffamily\Large\bfseries}
  {}{0pt}{}
\titleformat{\subsection}
  {\normalfont\sffamily\large\bfseries}
  {}{0pt}{}
\titleformat{\subsubsection}
  {\normalfont\sffamily\bfseries}
  {}{0pt}{}
\titleformat*{\paragraph}
  {\sffamily\normalsize}

\usepackage{fancyhdr}
\makeatletter
\let\ps@plain\ps@fancy
\fancyheadoffset[L]{4.5cm}
\fancyfootoffset[L]{4.5cm}

\definecolor{linky}{rgb}{0.0, 0.5, 1.0}

\newtcolorbox{repobox}
   {colback=red, colframe=red!75!black,
     boxrule=0.5pt, arc=2pt, left=6pt, right=6pt, top=3pt, bottom=3pt}

\newcommand{\ExternalLink}{%
   \tikz[x=1.2ex, y=1.2ex, baseline=-0.05ex]{%
       \begin{scope}[x=1ex, y=1ex]
           \clip (-0.1,-0.1)
               --++ (-0, 1.2)
               --++ (0.6, 0)
               --++ (0, -0.6)
               --++ (0.6, 0)
               --++ (0, -1);
           \path[draw,
               line width = 0.5,
               rounded corners=0.5]
               (0,0) rectangle (1,1);
       \end{scope}
       \path[draw, line width = 0.5] (0.5, 0.5)
           -- (1, 1);
       \path[draw, line width = 0.5] (0.6, 1)
           -- (1, 1) -- (1, 0.6);
       }
   }

\patchcmd{\@maketitle}{center}{flushleft}{}{}
\patchcmd{\@maketitle}{center}{flushleft}{}{}
\patchcmd{\@maketitle}{\LARGE}{\LARGE\sffamily}{}{}
\def\maketitle{{%
  
  \AB@maketitle}}
\makeatletter
\renewcommand\AB@affilsepx{ \protect\Affilfont}
\renewcommand\AB@affilnote[1]{{\bfseries #1}\hspace{3pt}}
\renewcommand{\affil}[2][]%
   {\newaffiltrue\let\AB@blk@and\AB@pand
      \if\relax#1\relax\def\AB@note{\AB@thenote}\else\def\AB@note{#1}%
        \setcounter{Maxaffil}{0}\fi
        \begingroup
        \let\href=\href@Orig
        \let\texttt=\textttOrig
        \let\protect\@unexpandable@protect
        \def\thanks{\protect\thanks}\def\footnote{\protect\footnote}%
        \@temptokena=\expandafter{\AB@authors}%
        {\def\\{\protect\\\protect\Affilfont}\xdef\AB@temp{#2}}%
         \xdef\AB@authors{\the\@temptokena\AB@las\AB@au@str
         \protect\\[\affilsep]\protect\Affilfont\AB@temp}%
         \gdef\AB@las{}\gdef\AB@au@str{}%
        {\def\\{, \ignorespaces}\xdef\AB@temp{#2}}%
        \@temptokena=\expandafter{\AB@affillist}%
        \xdef\AB@affillist{\the\@temptokena \AB@affilsep
          \AB@affilnote{\AB@note}\protect\Affilfont\AB@temp}%
      \endgroup
       \let\AB@affilsep\AB@affilsepx
}
\makeatother

\renewcommand\Affilfont{\sffamily\small\mdseries}
\ifnum 0\ifxetex 1\fi\ifluatex 1\fi=0 
  \usepackage[T1]{fontenc}
  \usepackage[utf8]{inputenc}

\else 
  \ifxetex
    \usepackage{mathspec}
    \usepackage{fontspec}

  \else
    \usepackage{fontspec}
  \fi
  \defaultfontfeatures{Ligatures=TeX,Scale=MatchLowercase}

\fi
\IfFileExists{upquote.sty}{\usepackage{upquote}}{}
\IfFileExists{microtype.sty}{%
\usepackage{microtype}
\UseMicrotypeSet[protrusion]{basicmath} 
}{}

\usepackage{hyperref}
\hypersetup{unicode=true,
            pdftitle={kuibit: Analyzing Einstein Toolkit simulations with Python},
            pdfborder={0 0 0},
            breaklinks=true}
\let\addcontentslineOrig=\addcontentsline
\def\addcontentsline#1#2#3{\bgroup
  \let\texttt=\textttOrig\addcontentslineOrig{#1}{#2}{#3}\egroup}
\let\markbothOrig\markboth
\def\markboth#1#2{\bgroup
  \let\texttt=\textttOrig\markbothOrig{#1}{#2}\egroup}
\let\markrightOrig\markright
\def\markright#1{\bgroup
  \let\texttt=\textttOrig\markrightOrig{#1}\egroup}

\usepackage{graphicx,grffile}
\makeatletter
\def\maxwidth{\ifdim\Gin@nat@width>\linewidth\linewidth\else\Gin@nat@width\fi}
\def\maxheight{\ifdim\Gin@nat@height>\textheight\textheight\else\Gin@nat@height\fi}
\makeatother
\setkeys{Gin}{width=\maxwidth,height=\maxheight,keepaspectratio}
\IfFileExists{parskip.sty}{%
\usepackage{parskip}
}{
\setlength{\parindent}{0pt}
\setlength{\parskip}{6pt plus 2pt minus 1pt}
}
\ifx\paragraph\undefined\else
\let\oldparagraph\paragraph
\renewcommand{\paragraph}[1]{\oldparagraph{#1}\mbox{}}
\fi
\ifx\subparagraph\undefined\else
\let\oldsubparagraph\subparagraph
\renewcommand{\subparagraph}[1]{\oldsubparagraph{#1}\mbox{}}
\fi

\title{kuibit: Analyzing Einstein Toolkit simulations with Python}

        \author[1]{Gabriele Bozzola}
    
      \affil[1]{Steward Observatory and Astronomy Department, The University
of Arizona}
  \date{\vspace{-7ex}}

\begin{document}
\maketitle

\marginpar{

  \begin{flushleft}
  \sffamily\small

  {\bfseries DOI:} \href{https://doi.org/10.21105/joss.03099}{\color{linky}{10.21105/joss.03099}}

  \vspace{2mm}

  {\bfseries Software}
  \begin{itemize}
    \setlength\itemsep{0em}
    \item \href{https://github.com/openjournals/joss-reviews/issues/3099}{\color{linky}{Review}} \ExternalLink
    \item \href{https://github.com/Sbozzolo/kuibit}{\color{linky}{Repository}} \ExternalLink
    \item \href{http://dx.doi.org/10.5281/zenodo.4681119}{\color{linky}{Archive}} \ExternalLink
  \end{itemize}

  \vspace{2mm}

  \par\noindent\hrulefill\par

  \vspace{2mm}

  {\bfseries Editor:} \href{https://researcher.watson.ibm.com/researcher/view.php?person=ibm-Eloisa.Bentivegna}{Eloisa Bentivegna} \ExternalLink \\
  \vspace{1mm}
    {\bfseries Reviewers:}
  \begin{itemize}
  \setlength\itemsep{0em}
    \item \href{https://github.com/yurlungur}{@yurlungur}
    \item \href{https://github.com/eloisabentivegna}{@eloisabentivegna}
    \end{itemize}
    \vspace{2mm}

  {\bfseries Submitted:} 30 January 2021\\
  {\bfseries Published:} 12 April 2021

  \vspace{2mm}
  {\bfseries License}\\
  Authors of papers retain copyright and release the work under a Creative Commons Attribution 4.0 International License (\href{http://creativecommons.org/licenses/by/4.0/}{\color{linky}{CC BY 4.0}}).

  \end{flushleft}
}

\hypertarget{summary}{%
\section{Summary}\label{summary}}

\texttt{kuibit}\footnote{A kuibit (harvest pole) is the tool traditionally used
  by the Tohono O'odham people to reach the fruit of the Saguaro cacti during
  the harvesting season.} is a Python library for analyzing simulations
performed with the \texttt{Einstein\ Toolkit}\footnote{While \texttt{kuibit} is
  designed for the \texttt{Einstein\ Toolkit}, most of its capabilities will
  work also for all the other codes based on \texttt{Cactus}~\cite{cactuscode}.
  For instance, it is known that \texttt{kuibit} can be used to analyze
  \texttt{Illinois\ GRMHD}~\cite{illinoisgrmhd}
  simulations.}~\cite{einsteintoolkit, einsteintoolkit2}, a free and open-source
code for numerical relativity and relativistic astrophysics. Over the past
years, numerical simulations like the ones enabled by the \texttt{Einstein\
  Toolkit} have become a critical tool in modeling, predicting, and
understanding several astrophysical phenomena, including binary black hole or
neutron star mergers. As a result of the recent detections of gravitational
waves by the LIGO-Virgo collaboration, these studies are at the forefront of
scientific research. The package presented in this paper, \texttt{kuibit},
provides an intuitive infrastructure to read and represent the output of the
\texttt{Einstein\ Toolkit}. This simplifies analyzing simulations and
significantly lowers the barrier in learning how to use the tool.

\hypertarget{statement-of-need}{%
\section{Statement of need}\label{statement-of-need}}

The \texttt{Einstein\ Toolkit} is a software for numerical simulations based on
the \texttt{Cactus}~\cite{cactuscode} computational framework and designed to be
accessible for both users and developers. Numerical-relativity simulations
require large and complex codes, which have to run on the world's largest
supercomputers. \texttt{Einstein\ Toolkit} significantly reduces this complexity
and improves accessibility by splitting infrastructure code from physics one. On
one side, there is memory management, parallelization, grid operations, and all
the other low-level details that are needed to successfully perform a simulation
but do not strictly depend on the physical system under consideration. On the
other, there are the physics modules, which implement the scientific aspects of
the simulation. Codes are developed by domain-experts and researchers can focus
on their goals without having to worry about the technical details of the
implementation. This makes the \texttt{Einstein\ Toolkit} easier to use and
extend.

Despite the advancements made by the \texttt{Einstein\ Toolkit}, there
is still a big leap between running a simulation and obtaining
scientific results. The output from the \texttt{Einstein\ Toolkit} is a
collection of files with different formats and structures, with data
that is typically spread across multiple files (one or more for each MPI
process) in various directories (one per checkpoint). Reading the
simulation output and properly combining all the data is a challenging
task. Even once the output is read, traditional data structures are not
a good representation of the physical quantities. For instance,
representing variables defined on an adaptive-mesh-refined grid as
simple arrays completely ignores all the information on the grid
structure, making some operations impractical or impossible to perform.
The lack of suitable interfaces introduces significant friction in
exploring the scientific content of a simulation. \texttt{kuibit} takes
care of both the aspects of reading the simulation data and of providing
high-level representations of the data that closely follows what
researchers are used to. In addition to this, \texttt{kuibit} also
includes a set of routines that are commonly used in the field: for
example, it handles unit conversion (including from geometrized units to
physical), it has the noise curves of known detectors, or it computes
gravitational-waves from simulation data.

\texttt{kuibit} is based on the same design (and in various cases,
implementation details too) of a pre-existing package named
\texttt{PostCactus}~\cite{pycactus}. Like \texttt{PostCactus},
\texttt{kuibit} has two groups of modules. The first is to define custom
data-types for time series, Fourier spectra, multipolar decompositions,
and grid data (both on uniform grids and mesh-refined ones). The second
group consists of the readers, which are a collection of tools to scan
the simulation output and organize it. The main reader is a class
\texttt{SimDir} which provides the interface to access all the data in
the simulation. For instance, the \texttt{timeseries} attribute in
\texttt{SimDir} is a dictionary-like object that contains all the time
series in the output. When reading data, \texttt{kuibit} takes care of
all the low-level details, like handling transparently simulation
restarts, or merging grid data stored in different files. Therefore,
users can easily access the data regardless of how complicated the
structure of the output is. Moreover, \texttt{kuibit} does not assume
any particular organization of the output and uses regular expressions
to find the relevant information from filenames or metadata, allowing
for flexibility in the simulation workflow.

Currently, \texttt{kuibit} is the only available package for quantitative
analysis of simulations that is free to use and that comes with documentation,
tutorials, and examples. Tools like \texttt{VisIt}~\cite{visit} or
\texttt{rugutils}~\cite{rugutils} focus only on visualizing grid data, while
other packages like \texttt{POWER}~\cite{POWER}, or
\texttt{pyGWAnalysis}~\cite{pyGWAnalysis} only on gravitational-wave data.
Capabilities similar to those of \texttt{kuibit} are offered by
\texttt{SimulationTools}~\cite{simulationtools}, that runs on the proprietary
Wolfram Mathematica, and by \texttt{PostCactus}~\cite{pycactus} and
\texttt{scidata}~\cite{scidata}, which, at the moment, do not support Python3
and do not have documentation. In addition to this, several research groups
develop their own private analysis software.

\texttt{kuibit} embraces the core principles of the \texttt{Einstein\ Toolkit}:
On one side, \texttt{kuibit} solves the engineering problems of reading and
representing \texttt{Einstein\ Toolkit} data so that researchers can directly
pursue their scientific goals without having to worry about how the data is
stored. With \texttt{kuibit}, the entry barrier into using the \texttt{Einstein\
  Toolkit} is the lowest it has ever been, and students and researchers can
inspect and visualize simulations in just a few lines of code. On the other
side, \texttt{kuibit} is designed to be a code for the community: it is free and
does not require any proprietary software to run, it is openly developed with an
emphasis on readability and maintainability, and it encourages contributions.

\hypertarget{acknowledgments}{%
\section{Acknowledgments}\label{acknowledgments}}

Gabriele Bozzola is supported by by the Frontera Fellowship by the Texas
Advanced Computing Center (TACC). Frontera~\cite{Frontera2020} is
founded by NSF grant OAC-1818253. This work was in part supported by NSF
Grant PHY-1912619 to the University of Arizona and made use of
computational resources provided by the Extreme Science and Engineering
Discovery Environment (XSEDE) under grant number TG-PHY190020. XSEDE is
supported by the NSF grant No.~ACI-1548562. Gabriele Bozzola wishes to
thank Wolfgang Kastaun for publicly releasing his \texttt{PostCactus}
package~\cite{pycactus} without which, \texttt{kuibit} would not exist.

\printbibliography

\end{document}